\documentclass[preprints,article,accept,oneauthor,pdftex]{Definitions/mdpi} 

\firstpage{1} 
\pubvolume{xx}
\issuenum{1}
\articlenumber{5}
\pubyear{2019}
\copyrightyear{2019}
\history{Received: 12 April 2019.}

\pdfoutput=1

\Title{Heavy-flavor measurements with the ALICE experiment at the LHC}

\Author{R\'obert V\'ertesi\orcidA{} (for the ALICE Collaboration)}

\AuthorNames{R\'obert V\'ertesi}

\address{\quad Wigner Research Centre for Physics
	\newline Current address: 1121 Budapest, Konkoly-Thege Mikl\'os \'ut 29-33., Hungary
	\newline Correspondence: vertesi.robert@wigner.mta.hu
}

\abstract{Heavy quarks (charm and beauty) are produced early in the nucleus-nucleus collisions, and heavy flavor survives throughout the later stages. 
Measurements of heavy-flavor quarks thus provide us with means to understand the properties of the Quark-Gluon Plasma, a hot and dense state of matter created in heavy-ion collisions.
Production of heavy-flavor in small collision systems, on the other hand, can be used to test Quantum-chromodynamics models. After a successful completion of the Run-I data taking period, the increased luminosity from the LHC and an upgraded ALICE detector system in the Run-II data taking period allows for an unprecedented precision in the study of heavy quarks. In this article we give an overview of selected recent results on heavy-flavor measurements with ALICE experiment at the LHC.}


\begin{document}


\section{Introduction}

High-energy heavy-ion collisions can recreate a strongly coupled Quark-Gluon Plasma (QGP), an extremely hot and dense, strongly interacting state of matter that was present in the early stages of the Universe. Smaller colliding systems such as pp or p-Pb are useful to study vacuum Quantum chromodynamics (QCD) and cold nuclear effects. Heavy quarks are produced early in the reaction and their numbers are almost conserved throughout the reaction. They undergo negligible flavor changing, and also there is very little thermal production or destruction within the QGP or the hadronic nuclear matter. Yet they are transported through the whole system, thus their kinematics can reveal transport properties such as collisional and radiational energy loss within the hot medium~\cite{Andronic:2015wma}. Since heavy flavor is preserved, it can be used as a penetrating probe down to very low momentum ($p \approx$0). Observing the hadronization of heavy-quarks reveal coalescence mechanisms in the hot medium, as well as fragmentation properties. This latter one, in comparison with light probes, is tell-tale about color charge and mass/flavor effects.

Measurements of heavy-flavor in pp collisions serve the primary purpose of setting a benchmark on their production and behavior in QCD vacuum. As in the case of many other probes, we rely on heavy-flavor production in pp collisions as a reference for measurements in larger collision systems where a substantial effect is expected by the hot or cold nuclear matter. In high-multiplicity events, however, we observe signatures of collectivity, such as long-range correlations~\cite{Khachatryan:2010gv}, that resemble those characteristic to QGP production. These are usually attributed to soft and semi-hard vacuum QCD effects such as multiple-parton interactions (MPI)~\cite{Bartalini:2011jp}, although alternative explanations also exist~\cite{Schenke:2017bog}. Heavy-flavor measurements versus event activity play a key role in clarifying the origin of such signatures. Fragmentation of heavy quarks can be studied using jet and correlation observables. Besides color charge, mass and flavor effects these can pin down the contribution of late gluon splitting to heavy-flavor production.
Production of the heavy-flavored baryons provides additional information on heavy-flavor fragmentation and provides key information for the development of theoretical models.

Some of the most interesting recent heavy-flavor results  carried out by the ALICE experiment are summarized in the following sections. The details of the ALICE detector system are described elsewhere~\cite{Abelev:2014ffa}. Heavy-flavor is accessed either directly, via the full reconstruction of the decay products, or indirectly by measuring some of the decay products (decay leptons for most of the cases) and statistically disentangling heavy-flavor contribution. 

\section{Heavy-flavor mesons in the QCD vacuum}

\begin{figure}[h]
	\centering
	\includegraphics[width=0.5\columnwidth]{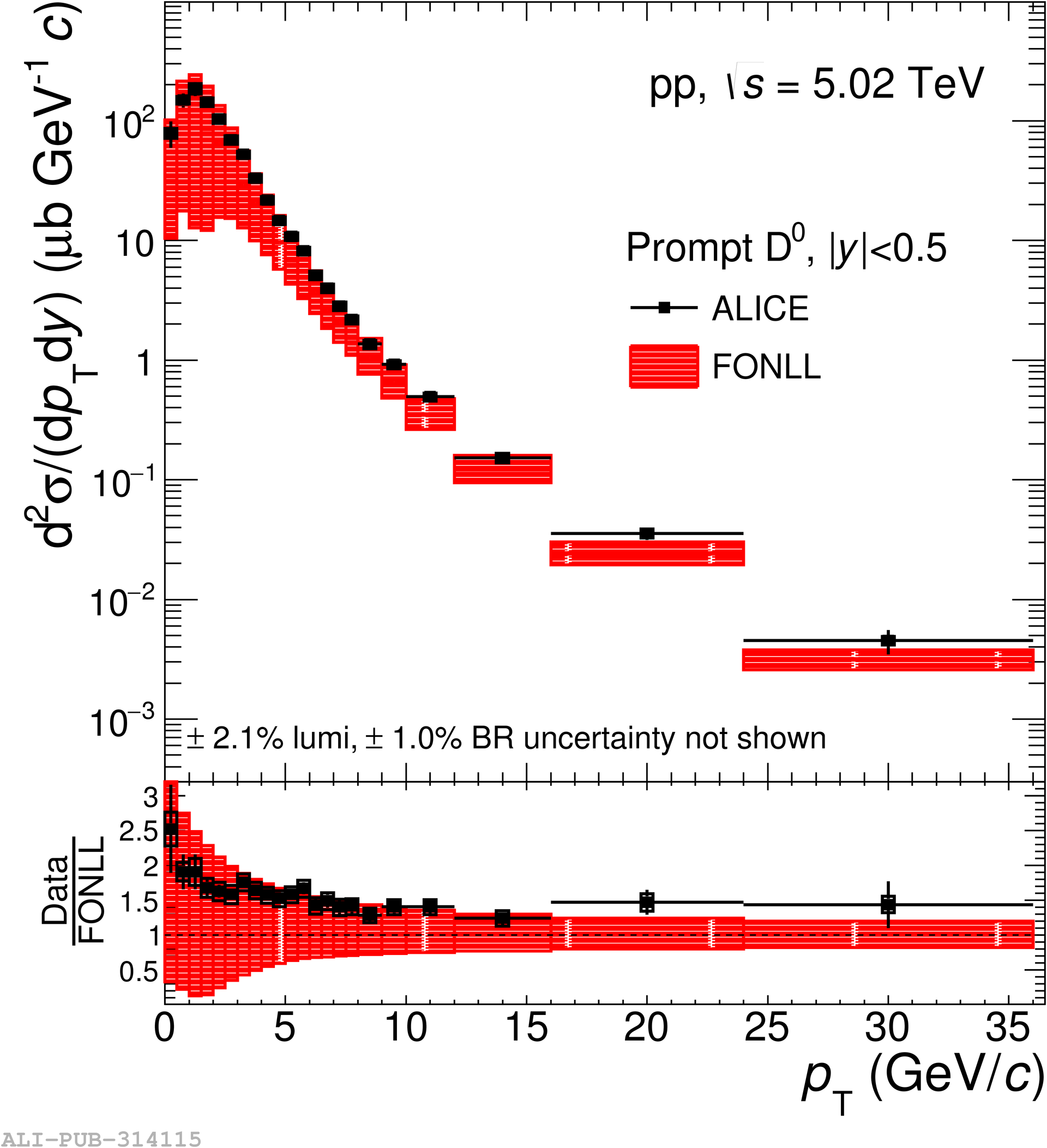}
	\hspace{0.05\columnwidth}
	\includegraphics[width=0.40\columnwidth]{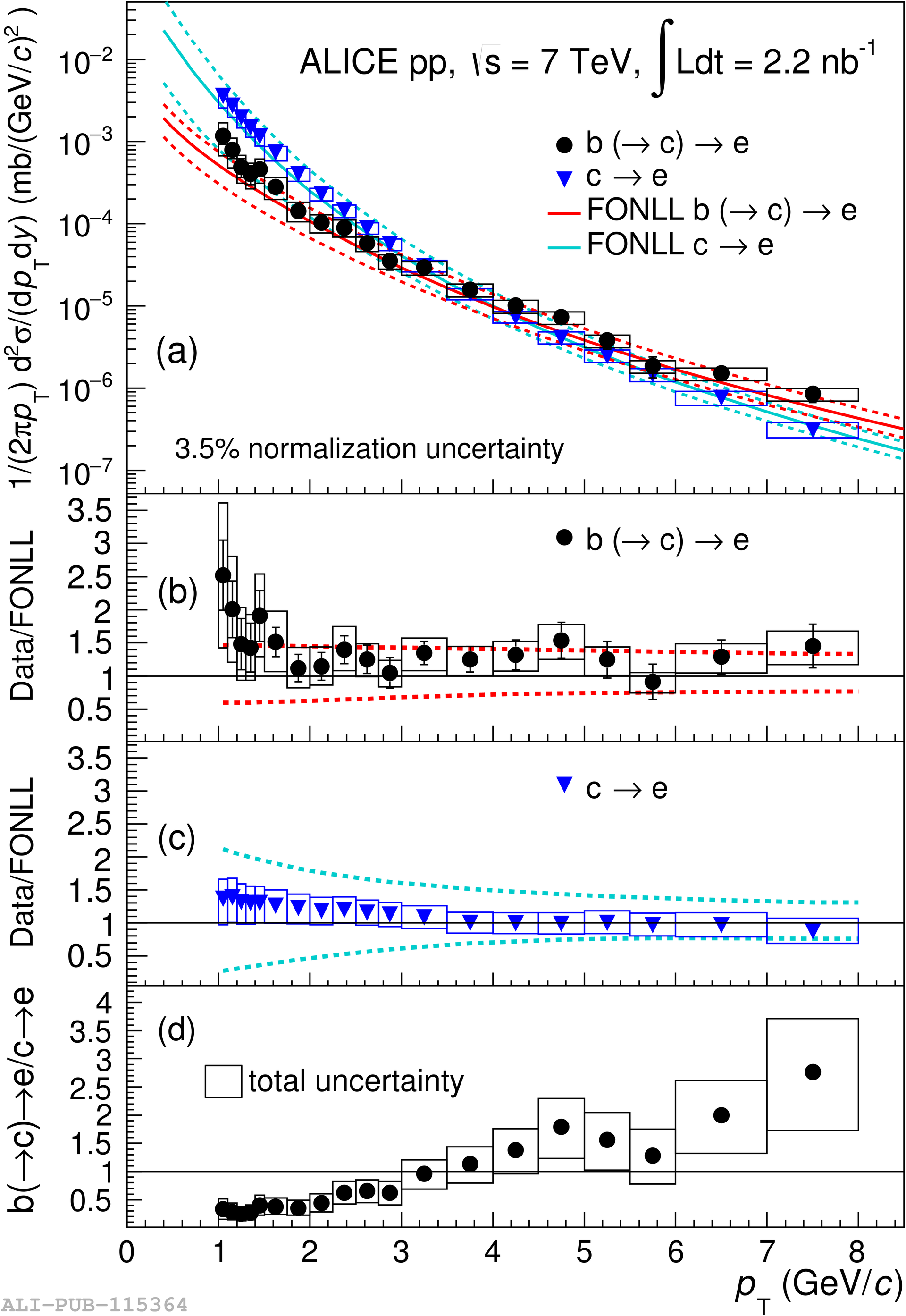}
	\caption{\label{fig:xsec}{\bf(Left)} Production cross section of $D^0$ mesons in pp collisions at $\sqrt{s_{NN}}=5.02$ TeV, compared to a theoretical calculation~\cite{Cacciari:2012ny}. {\bf(Right)} Production cross section and beauty faction of heavy-flavor electrons in pp collisions at $\sqrt{s_{NN}}=7$ TeV, compared to a theoretical calculation~\cite{Cacciari:2012ny}.}
\end{figure}

Heavy flavour mesons in pp collisions help understand the physics of QCD vacuum. Direct and indirect measurements on their production are carried out in ALICE.
While direct reconstruction of charmed mesons such as $D^0$, $D^+$ and $D^{*+}$ provide a more direct access to decay kinematics, indirect measurements through semi-leptonic decays provide a mixture of beauty and charm contributions. In recent measurements, however, secondary vertexing with fine resolution in the Inner Tracking System (ITS) provide means to statistically separate the contributions of charm and beauty decays.
Fig.~\ref{fig:xsec} shows recent measurements of $D^0$ mesons as well as heavy-flavor decay electrons. In general, several theoretical models describe the measurements within uncertainties, and heavy-flavor meson measurements already provide restrictive input to them.

\begin{figure}[h]
	\centering
	\includegraphics[width=0.40\columnwidth]{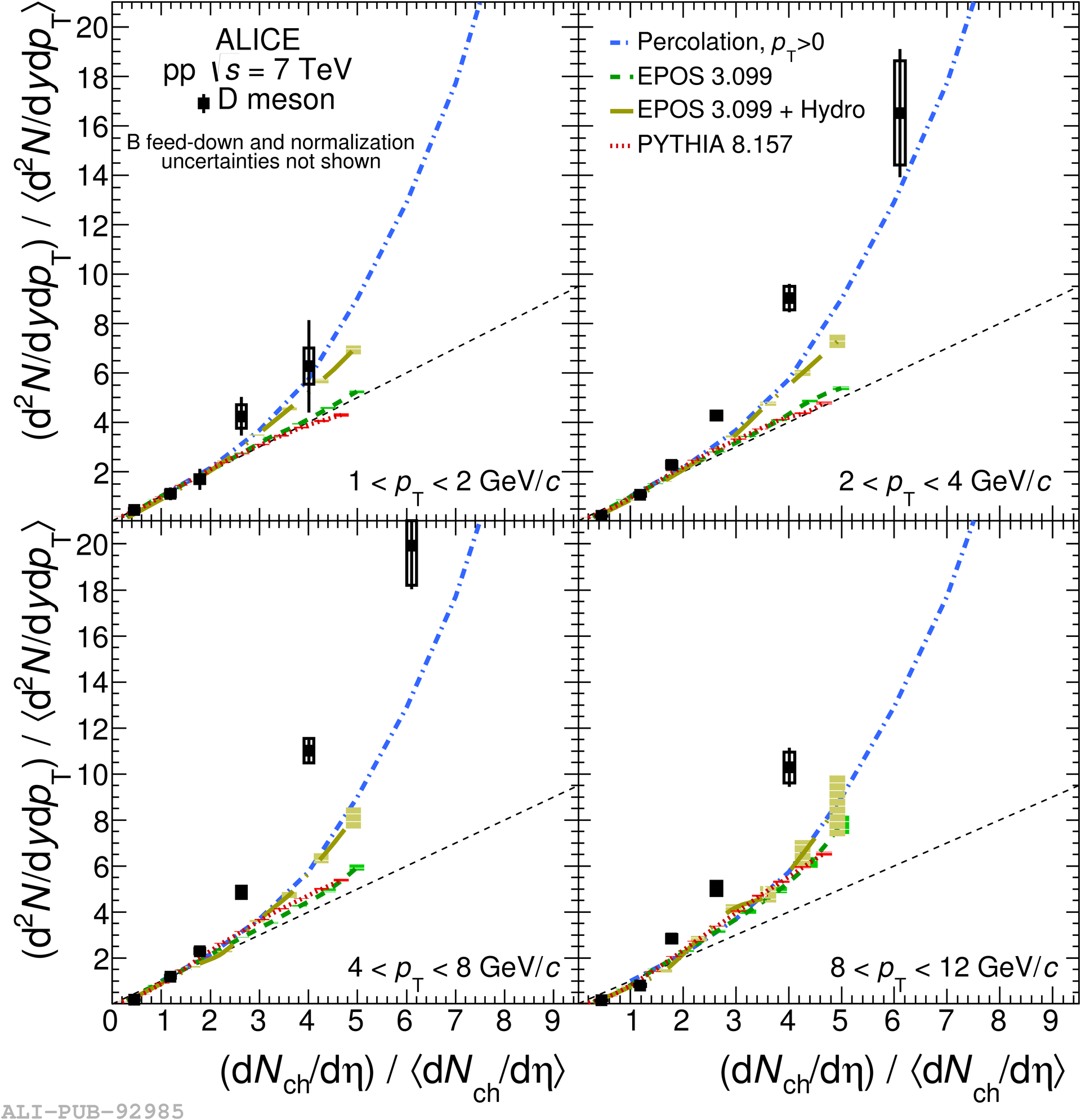}
	\includegraphics[width=0.59\columnwidth]{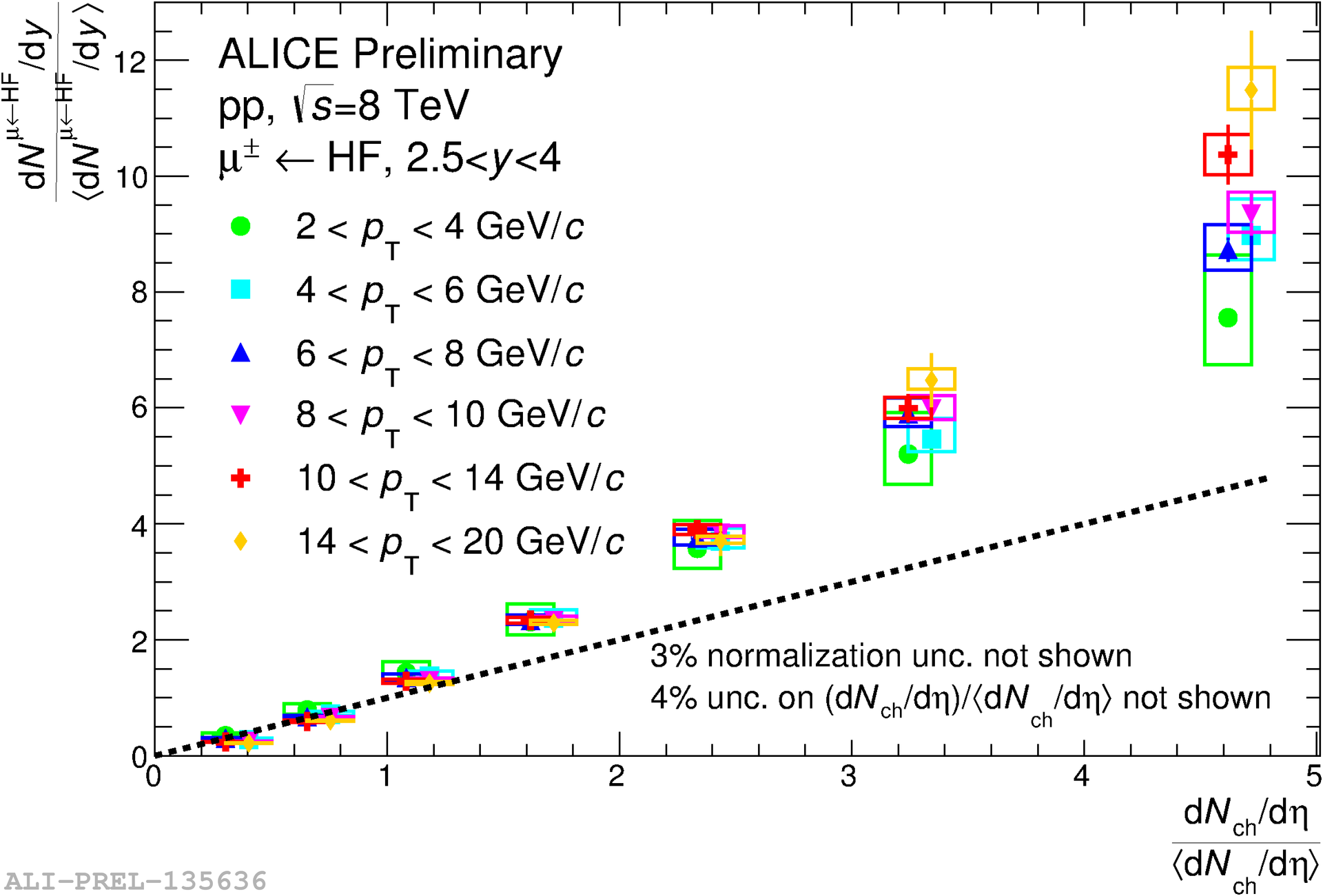}
	\caption{\label{fig:mult}\textbf{(Left)} Average self-normalized yields of D mesons in pp collisions at $\sqrt{s_{NN}}=7$ TeV at mid-rapidity~\cite{Adam:2015ota} compared to several model calculations~\cite{Sjostrand:2007gs,Ferreiro:2015gea,Werner:2013tya}, and \textbf{(right)} of muons in pp collisions at $\sqrt{s_{NN}}=8$ TeV at forward rapidity, for several transverse momentum ranges.}
\end{figure}

In pp collisions, the relative yield of D-mesons at mid-rapidity depends steeper than linearly on the relative charged multiplicity~\cite{Adam:2015ota}. This shows that hard processes such as heavy-flavor production and soft processes like bulk charged hadron production scale differently with event activity. Comparison to theoretical calculations~\cite{Sjostrand:2007gs,Ferreiro:2015gea,Werner:2013tya} in the left panel of Fig.~\ref{fig:mult} shows that the steeper-than-linear trend can be qualitatively described by calculations that include multiple parton interactions (MPI). A recent measurement with heavy-flavor muons indicates a similar trend at forward rapidity, as shown in the right panel of Fig.~\ref{fig:mult}.

\section{The baryonic sector}

\begin{figure}[H]
	\centering
	\includegraphics[width=0.4\columnwidth]{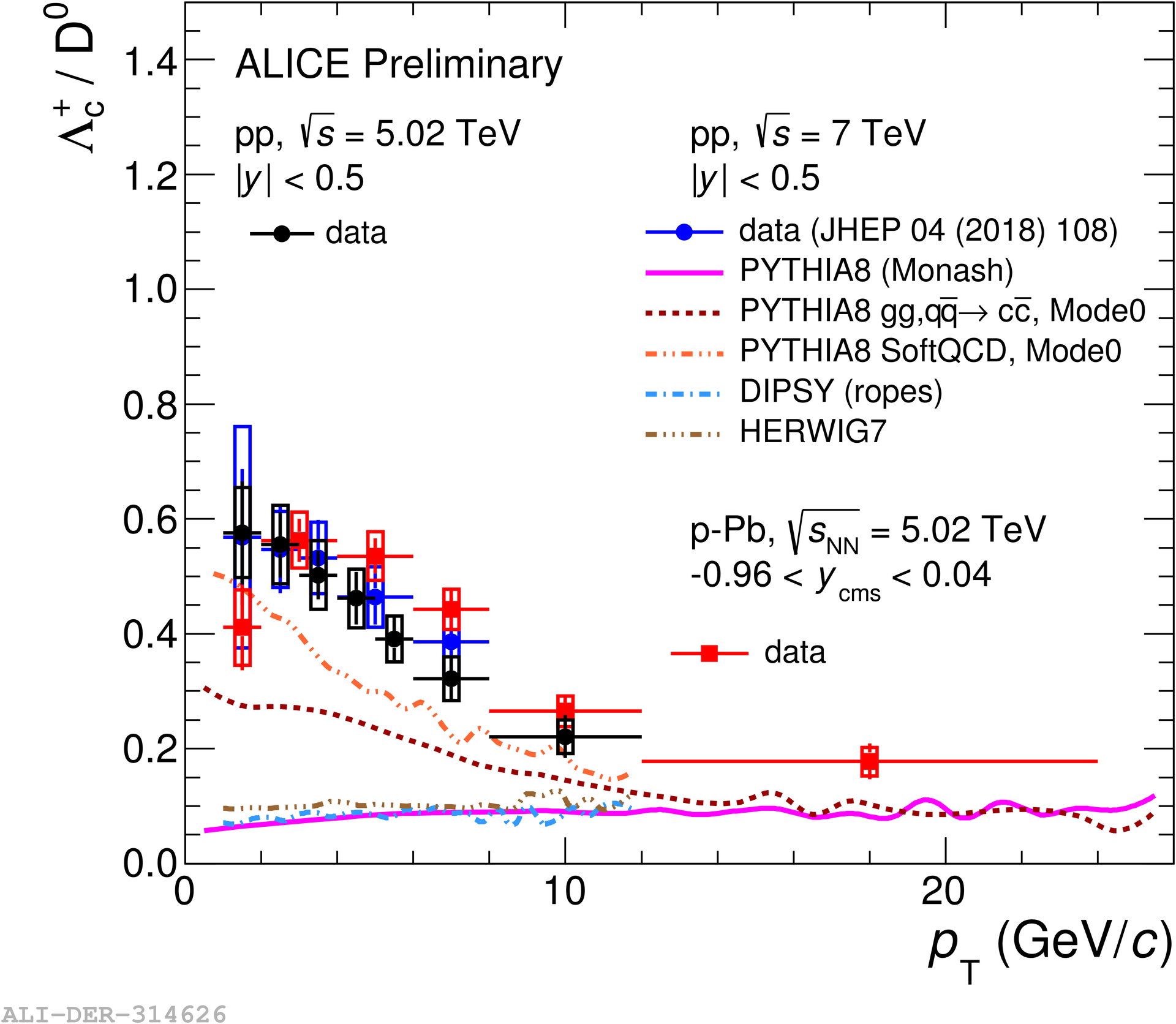}
	\includegraphics[width=0.5\columnwidth]{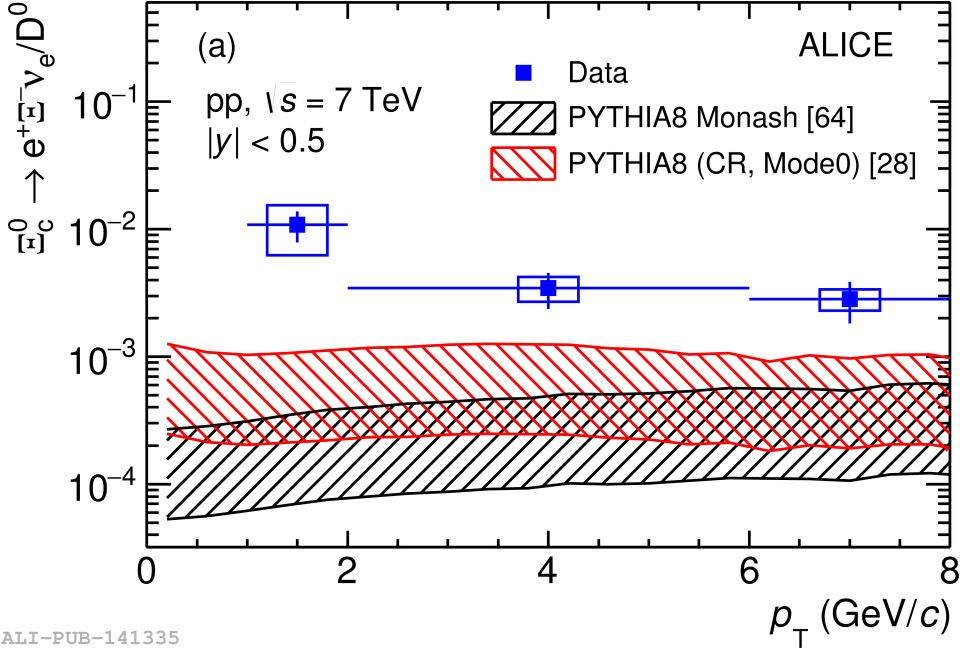}
	\caption{\label{fig:baryon}\textbf{(Left)} Ratios of $\Lambda_c^+$ and \textbf{(right)}  $\Xi_c^0$ $p_\mathrm{T}$-differential cross sections~\cite{Acharya:2017lwf} over $D^0$ in pp collisions at $\sqrt{s}=7$ TeV, compared to several theoretical calculations~\cite{Cacciari:2012ny,Bahr:2008pv,Bierlich:2015rha}.}
\end{figure}

Measurements of baryons with charm content provide valuable input for theories to understand heavy-flavor fragmentation. Recent measurements of $\Lambda_c^+$ in pp collisions at $\sqrt{s}=5$ and 7 TeV and LHC-first $\Xi_c^0$ measurements in pp collisions at $\sqrt{s}=7$ TeV~\cite{Acharya:2017lwf} show that the production of these mesons are underestimated by widely used theoretical models~\cite{Cacciari:2012ny,Bahr:2008pv,Bierlich:2015rha}. The same is observed in charmed baryon-to-meson ratios with a decreased relative uncertainty, as shown on Fig.~\ref{fig:baryon}. This shows that our current understanding on heavy-flavor fragmentation in the baryon sector is inadequate.

\section{Heavy-flavor in cold nuclear matter}

Cold nuclear matter (CNM) effects are expected to appear both in the initial and in the final state in collisions of protons on heavy ions, in an environment of substantial volume where quarks are still confined into hadrons. Several models predict modification of the nuclear parton distribution functions (nPDF) by (anti)shadowing, gluon saturation. A noon-negligible energy loss in the CNM is also expected, as well as the transverse-momentum ($k_T$) broadening of the initial and final state partons~\cite{Beraudo:2015wsd,Fujii:2013yja,Mangano:1991jk,Vitev:2007ve,Kang:2014hha}.
While collectivity has also been observed in p-A collisions, the question whether deconfined matter can be created in p-A collisions has not yet been settled~\cite{Schenke:2017bog}.

\begin{figure}[h]
	\centering
	\includegraphics[width=0.42\columnwidth]{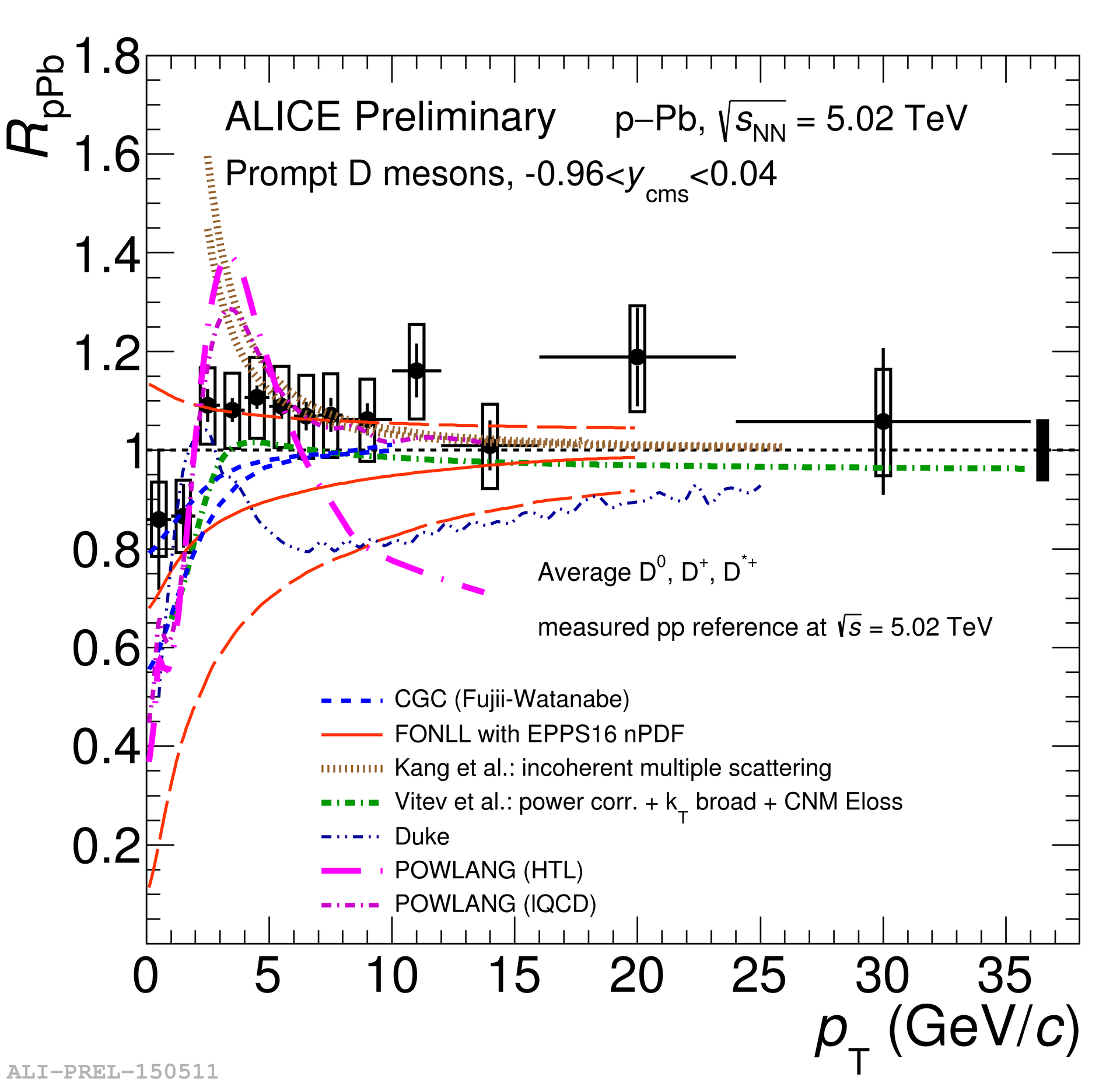}
	\includegraphics[width=0.4\columnwidth]{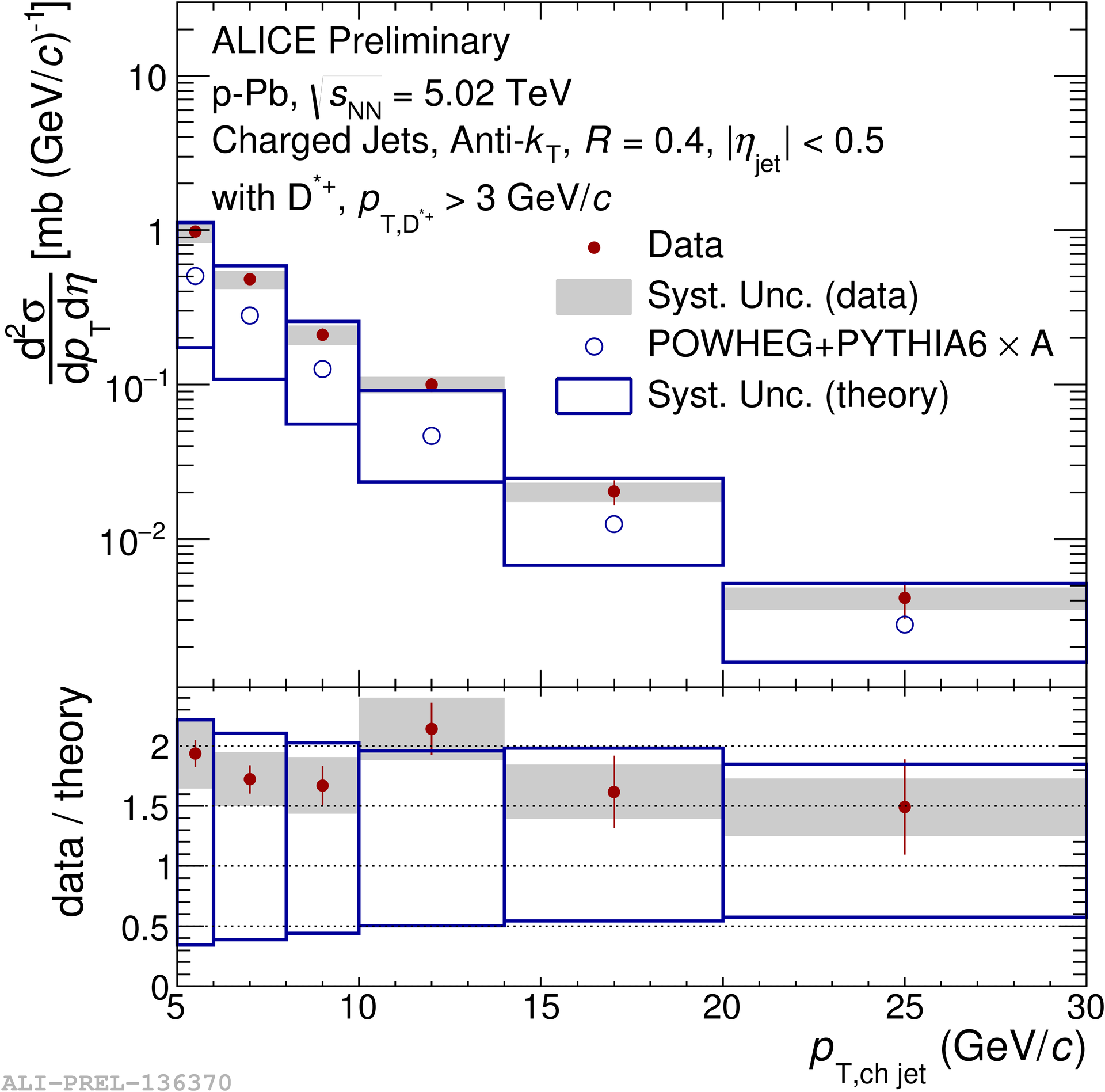}	\caption{\label{fig:DpA}(\textbf{Left}) Nuclear modification factor $R_{pPb}$ of prompt D mesons in p-Pb collisions at $\sqrt{s_{NN}}=5.02$ TeV, compared to several models~\cite{Beraudo:2015wsd,Fujii:2013yja,Mangano:1991jk,Vitev:2007ve,Kang:2014hha}.
	(\textbf{Right}.) Cross section of $D^{*+}$-tagged charged jets in p-Pb collisions at $\sqrt{s_{NN}}=5.02$ TeV, compared to pQCD NLO calculations~\cite{Nason:2004rx}.}
\end{figure}

The left panel of Fig.~\ref{fig:DpA} shows the nuclear modification factor $R_{pPb}$ of prompt D mesons as recently measured by the ALICE measurement in p-Pb collisions at $\sqrt{s_{NN}}=5.02$ TeV. 
The measurement extends down to $p_\mathrm{T}\approx 0$ and $R_{pPb}$ is consistent with unity throughout the range. Several models incorporating different CNM mechanisms~\cite{Fujii:2013yja,Mangano:1991jk,Vitev:2007ve,Kang:2014hha} adequately describe the weak nuclear modification. The POWLANG model with lattice-QCD calculations, which incorporates QGP formation in a small volume~\cite{Beraudo:2015wsd}, is also able to describe data in a statistically acceptable manner.
The right panel of Fig.~\ref{fig:DpA} shows charmed jet measurements, defined as jets containing D-mesons with $p_\mathrm{T}>3$ GeV/$c$ reconstructed within a jet. POWHEG pQCD NLO calculations with PYTHIA fragmentation~\cite{Nason:2004rx,Sjostrand:2007gs} describe data within uncertainties, indicating the lack of a strong nuclear modification of heavy-flavor jets. However, since the theoretical predictions have large uncertainties, the current measurements provide strong constraints for model development and tuning.

\section{Nuclear modification and collectivity in hot nuclear matter}

\begin{figure}[h]
	\centering
	\includegraphics[width=0.40\columnwidth]{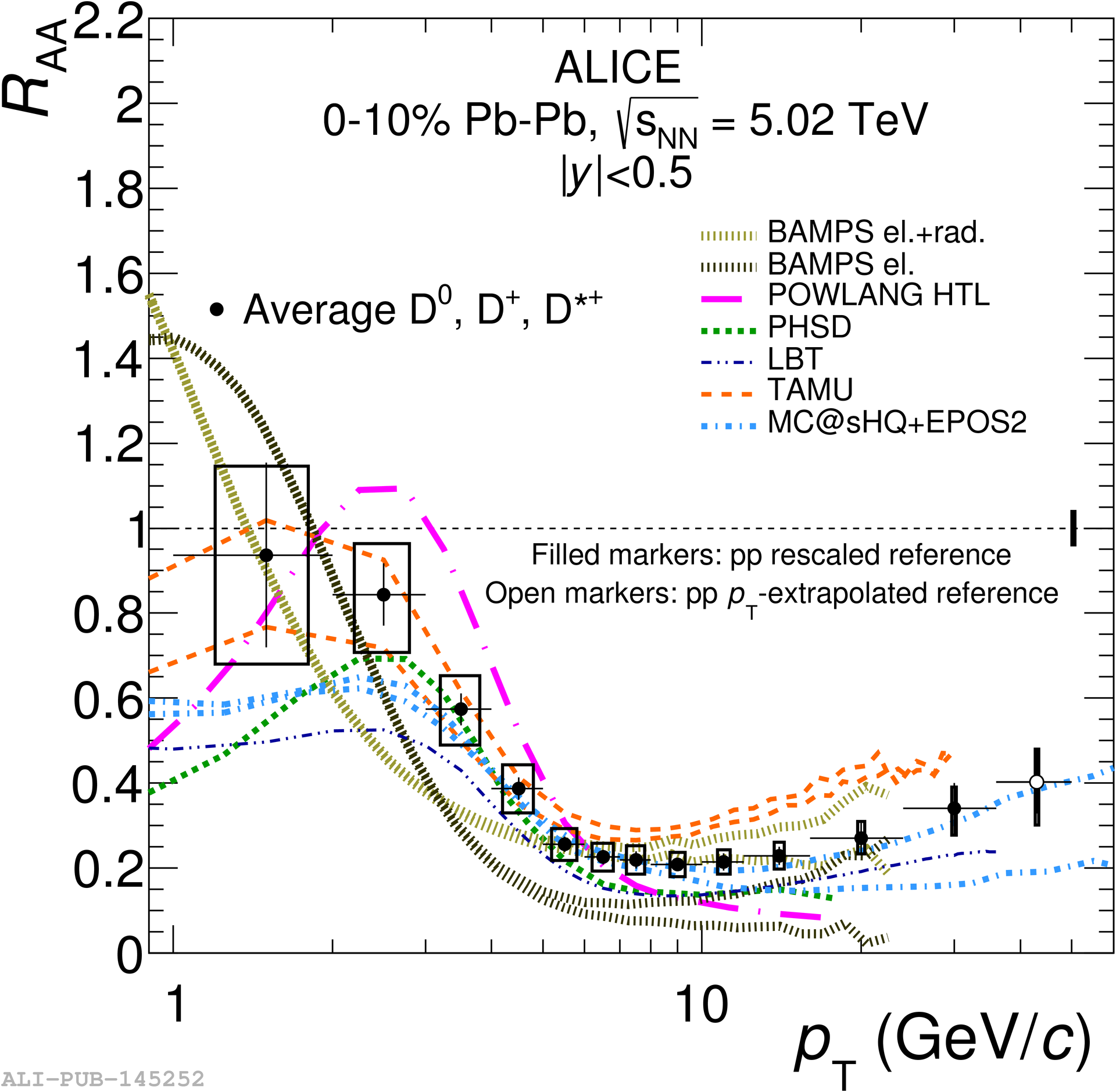}
	\includegraphics[width=0.58\columnwidth]{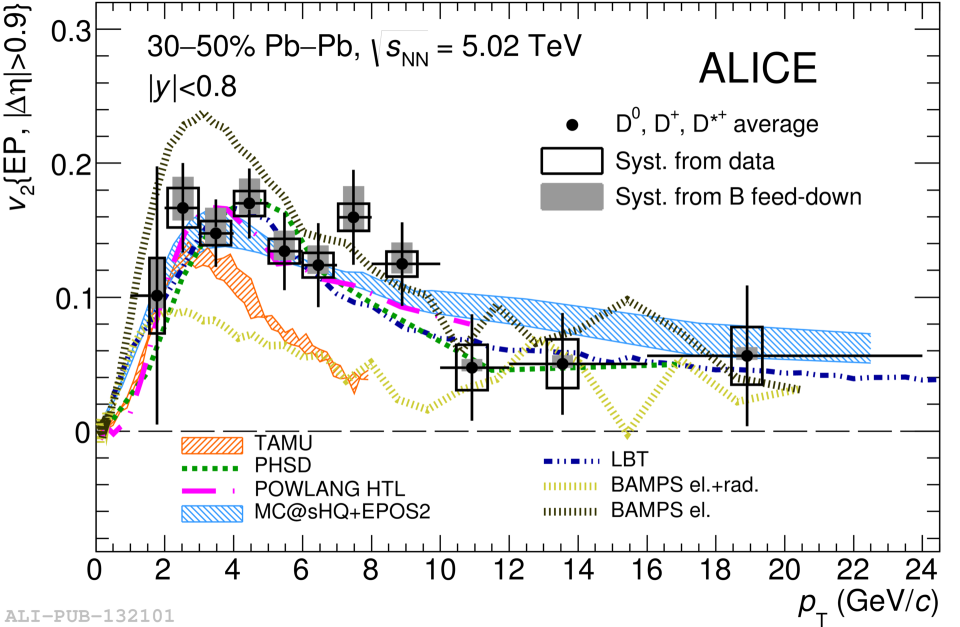}
	\caption{\label{fig:v2}(\textbf{Left}) Average nuclear modification factor $R_{AA}$ of 
	$D^0$, $D^+$ and $D^{*+}$ mesons in central Pb-Pb collisions at $\sqrt{s_\mathrm{NN}}=5.02$ TeV~\cite{Acharya:2018hre}, compared to transport model calculations~\cite{Averbeck:2011ga,Cao:2017hhk,He:2014cla,Nahrgang:2013xaa,Song:2015ykw}.
	(\textbf{Right}) Average azimuthal anisotropy $v_2$ of D mesons in semi-central Pb-Pb collisions at $\sqrt{s_\mathrm{NN}}=5.02$ TeV~\cite{Acharya:2017qps}, compared to model calculations~\cite{Averbeck:2011ga,Cao:2017hhk,He:2014cla,Nahrgang:2013xaa,Song:2015ykw}.}
\end{figure}

The nuclear modification factor $R_{AA}$ of heavy flavor in AA collisions is sensitive to radiative and collisional energy loss processes within the medium and can probe color charge effects as well as flavor-dependent hadronization. At higher momenta, little difference is found between $R_{AA}$ of charmed and light mesons, and both can be described by pQCD calculations~\cite{Acharya:2018hre}. Nuclear modification of heavy flavor at lower momenta, however, shows a significantly weaker suppression pattern than that of light flavor. The left panel of Fig.~\ref{fig:v2} shows the $R_{AA}$ of heavy flavor compared to several model calculations with  different ingredients regarding heavy flavor transport~\cite{Averbeck:2011ga,Cao:2017hhk,He:2014cla,Nahrgang:2013xaa,Song:2015ykw}. Models that contain charm-light coalescence~\cite{Cao:2017hhk,He:2014cla,Nahrgang:2013xaa,Song:2015ykw} typically provide better description of the dataset.

To achieve a stronger discriminative power of data over models, the azimuthal anisothropy parameter $v_2$ ("elliptic flow") of $D_0$ mesons in semi-central pp collisions is shown in Fig.~\ref{fig:v2}. A substantial heavy-flavor anisotropy can be observed. The $v_2$ of the $D_0$ mesons is qualitatively similar to that observed for light mesons ($\pi^\pm$) in Pb-Pb collisions at $\sqrt{s_\mathrm{NN}}=5.02$ TeV.

\section{Summary and outlook}

We gave an overview of selected recent heavy-flavor results from ALICE in pp, p-Pb and Pb-Pb colliding systems. Transverse momentum differential production of both the charmed and beauty mesons in pp collisions are generally described by pQCD models within uncertainties. The production of charmed baryons is, however, underestimated by theoretical calculations, indicating that models for fragmentation need improvement. The production of heavy flavor increases steeper-than-linearly with event activity, indicating the role of multiple-parton interactions. The models, however, fail to describe the data quantitatively. 
Nuclear modification by cold nuclear matter is weak in case of both the D mesons and reconstructed D-jets. These recent Run-2 measurements already provide strong restrictions for theoretical calculations. 
While the suppression of charmed D mesons is similar to that of light hadrons at high-$p_\mathrm{T}$, low-$p_\mathrm{T}$ suppression is weaker. A substantial azimuthal anisotropy can be observed for charmed mesons. Although the simultaneous description of  $R_{AA}$ and $v_2$ is a challenge for theory, some transport models that incorporate mechanisms for coalescence between charm quarks and light quarks adequately describe the low-$p_\mathrm{T}$ behavior of both observables.
Ongoing heavy-flavor measurements at ALICE show unprecedented precision down to very low momenta. The Run-3 phase of LHC with further increased luminosity and detector upgrades~\cite{Abelevetal:2014cna} will bring the era of precision beauty measurements.

\vspace{6pt} 

\funding{This work has been supported by the Hungarian NKFIH/OTKA K 120660 grant and the János Bolyai scholarship of the Hungarian Academy of Sciences.}


\conflictsofinterest{The author declares no conflict of interest.} 

\reftitle{References}



\sampleavailability{Published ALICE data are available in the HEPData repository.}


\end{document}